\documentclass[12pt,preprint]{aastex}

\newcommand  \Hubble   {\ifmmode {\rm km\,s}^{-1}\,{\rm Mpc}^{-1}
                        \else km\,s$^{-1}$\,Mpc$^{-1}$\fi}
\newcommand  \Msun     {\ifmmode M_{\odot} \else M$_{\odot}$\fi}
\newcommand  \Lsun     {\ifmmode L_{\odot} \else $L_{\odot}$\fi}
\newcommand  \cms      {\ifmmode {\rm cm\,s}^{-1} \else cm\,s$^{-1}$\fi}
\newcommand  \acc      {\ifmmode {\rm km\,s}^{-2} \else km\,s$^{-2}$\fi}
\newcommand  \kms      {\ifmmode {\rm km\,s}^{-1} \else km\,s$^{-1}$\fi}

\newcommand  \ergs     {\ifmmode {\rm erg\,s}^{-1} \else erg s$^{-1}$\fi}
\newcommand  \ergcms   {\ifmmode {\rm erg\,cm}^{-2}\,{\rm s}^{-1}
                        \else erg\,cm$^{-2}$\,s$^{-1}$\fi}
\newcommand  \ergcmsA  {\ifmmode{\rm erg\,cm}^{-2}\,{\rm s}^{-1}\,{\rm \AA}^{-1}
                        \else erg\,cm$^{-2}$\,s$^{-1}$\,\AA$^{-1}$\fi}
\newcommand  \ergcmsHz {\ifmmode{\rm erg\,cm}^{-2}\,{\rm s}^{-1}\,{\rm Hz}^{-1}
                        \else erg\,cm$^{-2}$\,s$^{-1}$\,Hz$^{-1}$\fi}
\newcommand  \phcms    {\ifmmode {\rm photons\,cm}^{-2}\,{\rm s}^{-1}
                        \else photons\,cm$^{-2}$\,s$^{-1}$\fi}
\newcommand  \phcmsA   {\ifmmode {\rm photons\,cm}^{-2}\,{\rm s}^{-1}\,{\rm\AA}^{-1}
                        \else photons\,cm$^{-2}$\,s$^{-1}$\,\AA$^{-1}$\fi}
\newcommand  \rblr     {R$_{\rm BLR}$}

\begin{document}

\title{Reverberation Mapping of the Intermediate Mass Nuclear Black Hole in SDSS J114008.71$+$030711.4}
\author{Stephen E. Rafter\altaffilmark{1}, Shai Kaspi\altaffilmark{1}, Ehud Behar\altaffilmark{1}, Wolfram Kollatschny\altaffilmark{2}, Matthias Zetzl\altaffilmark{2}}
\altaffiltext{1}{Physics Department, the Technion, Haifa 32000, Israel; e-mail: rafter, shai, behar: @physics.technion.ac.il}
\altaffiltext{2}{Institut f\"{u}r Astrophysik, Universit\"{a}t G\"{o}ttingen, Friedrich-Hund Platz 1, 37077 G\"{o}ttingen, Germany; email: wkollat@astro.physik.uni-goettingen.de}
\begin{abstract}
\indent

We present the results of a reverberation mapping (RM) campaign on the black hole (BH) associated with the active galactic nucleus (AGN) in SDSS J114008.71$+$030711.4 (hereafter GH08).  This object is selected from a sample of 19 candidate intermediate mass BHs (M$_{\rm BH} < 10^{6}$ M$_{\odot}$) found by \citet{2004ApJ...610..722G} in the Sloan Digital Sky Survey (SDSS).  We used the Hobby-Eberly Telescope to obtain 30 spectra over a period of 178 days in an attempt to resolve the reverberation time lag ($\tau$) between the continuum source and the broad line region (BLR) in order to determine the radius of the BLR (\rblr) in GH08.  We measure $\tau$ to be 2 days with an upper limit of 6 days.  We estimate the AGN luminosity at 5100 \AA\ to be $\lambda$L$_{\rm 5100} \approx 1.1 \times 10^{43}$ \ergs\ after deconvolution from the host galaxy.  The most well calibrated \rblr$-$L relation predicts a time lag which is 4 times larger than what we measure.  Using the measured H$\beta$ full-width-at-half-maximum of 703 $\pm$ 110 \kms\ and an upper limit for \rblr\ $= 6$ light days, we find M$_{\rm BH} \lesssim 5.8 \times 10^{5}$ \Msun\ as an upper limit to the BH virial mass in GH08, which implies super$-$Eddington accretion.  Based on our measured M$_{\rm BH}$ we propose that GH08 may be another candidate to add to the very short list of AGNs with M$_{\rm BH} < 10^{6}$ M$_{\odot}$ determined using RM.

\end{abstract}


~~~~~.

\section{Introduction}
\indent

The determination of fundamental physical parameters is paramount for a complete understanding of the mechanism powering Active Galactic Nuclei (AGNs) and their role in galaxy evolution.  The evidence for coevolution  of the host galaxy and the central massive black hole is found observationally by the well known M$-\sigma_{*}$ relation \citep{2000ApJ...539L...9F, 2002ApJ...574..740T} but it's applicability for low mass sources is still under debate \citep[see for e.g.,][]{2011Natur.469..374K, 2011arXiv1102.0537M}.  AGN come in a variety of flavors and are found to have a range in bolometric luminosity from as low as $10^{42}$ \ergs\ to beyond $10^{48}$ \ergs, and accordingly, have a large range in their central black hole mass (M$_{\rm BH}$) from around $10^{5}$ \Msun\ up to several $10^{9}$ \Msun .  For about 20 years the method which has yielded the most robust determinations of M$_{\rm BH}$ in AGN, over a range in luminosity and M$_{\rm BH}$, is reverberation mapping (RM).  The RM technique allows determination of the time-lag ($\tau$) between a change in continuum emission from an accretion disk and the delayed response due to light travel time of line emission from high velocity clouds in a distant broad-line region (BLR).  Combining measurements of the velocity range ($\Delta V$) from an emission line width originating in the BLR and the distance of the BLR from the BH (\rblr\ $= c\tau$, where c is the speed of light) gives the virial mass within the BLR.  Since the BH dominates the mass inside the BLR, M$_{\rm BH} = f\Delta$V$^{2}$\rblr$G^{-1}$ , where $f$ is a scaling factor that depends on the geometry, kinematics and orientation of the BLR and $G$ is the gravitational constant \citep[for a full review see][]{2004ApJ...613..682P}.  Assuming an isotropic BLR circular velocity field and that the full width at half maximum (FWHM) reflects two times the typical BLR velocity yields $f =$ 0.75 \citep{2000ApJ...533..631K}.  The most appropriate value of $f$ to use depends on the method used to find $\Delta$V and reported values vary from 0.5 up to 5.5 depending on the FWHM of the line and the FWHM/$\sigma_{line}$ ratio.  \citet{2006A&A...456...75C} define 2 populations based on the FWHM/$\sigma_{line}$ ratio being below and above 2.35 (the intrinsic ratio when fitting lines with a Gaussian) and estimate $f$ for each population to be 1.5 and 0.5 respectively.  Therefore a reasonable $f$ factor to use when measuring a line width measured from the FWHM of a Gaussian fit is 0.75.  When R$_{\rm BLR}$ is in light days and $\Delta$V measured by the FWHM in 1000 \kms, the BH mass in solar masses is:

\begin{equation}
M_{\rm BH} = 1.464 \times 10^{5}R_{\rm BLR}\Delta V^{2} \Msun
\end{equation}

When the RM determined \rblr\ is plotted against the AGN continuum luminosity a general trend is revealed where \rblr\ $\propto$ L$^{\alpha}$, the so called radius$-$luminosity (\rblr$-$L) power-law relation.  An important caveat of the \rblr$-$L relation is the difficulty to accurately determine the contribution from the inner region of the host galaxy to the luminosity of the AGN.  \citet{2006ApJ...644..133B} show, using high resolution Hubble Space Telescope (HST) images, that an accurate determination of the AGN luminosity with respect to the host galaxy luminosity can change the slope ($\alpha$) of this relation significantly.  A physical consequence of this relation is that for AGNs with very high luminosities, $\tau$ can be very large, on the order of several to tens of years.  Since the RM technique requires a high degree of accuracy in the calibration of spectra over an extended period of time for the most luminous AGNs, only a few of the highest luminosity, and therefore the highest mass AGNs have been attempted with RM \citep{2007ApJ...659..997K}.  On the other hand, for AGNs with the lowest luminosity, $\tau$ can be quite short, on the order of a few hours to a few days.  Again, in terms of observations RM turns out to be difficult.  If $\tau$ is on the order of hours, then semi-weekly observations will not resolve the time lag.  The largest group of AGNs with M$_{\rm BH}$ determined using RM is the nearby Seyfert galaxies \citep{1999ApJ...526..579W, 2010ApJ...716..993B, 2010ApJ...721..715D} and several PG quasars \citep{2000ApJ...533..631K} who have $\tau$ on the order of a few to several hundreds of days and $10^{6} <$ M$_{\rm BH}/$\Msun $\lesssim 10^{9}$.  AGNs with M$_{\rm BH} < 10^{6}$ \Msun\ will have small $\tau$, on the order of hours to days, and slower bulk motion of clouds in the BLR as determined by the width of the broad lines.  For example if $\tau\approx 7$ days then $\Delta V < 1000$ \kms\ for M$_{\rm BH} < 10^{6}$ \Msun\ .  Only a very few AGN meeting these criteria have had M$_{\rm BH}$ robustly determined using RM.  The AGN in NGC 4395 is one such object whose M$_{\rm BH} \approx 3.6 \times 10^{5}$ \Msun\ was determined by \citet{2005ApJ...632..799P} using HST UV spectra to find a C{\sc iv} time lag of about 1 hour.  \citet{2006ApJ...650...88D} tenuously measure a ground based Balmer line time lag for NGC 4395 that is consistent with the HST measurement.  NGC 4395 is notorious in that it does not have a classic bulge and therefore does not follow the M$- \sigma_{*}$ relation.  NGC 4051 was found by \citet{2003MNRAS.343.1341S} to host a NLS1$-$like AGN with M$_{\rm BH} \approx 5^{+6}_{-3} \times 10^{5}$ \Msun\ measured using RM.  \citet{2004IAUS..222...65W} subsequently show that NGC 4051 follows the M$- \sigma_{*}$ relation but has an underluminous bulge when compared to more standard type 1 AGN.  More recently \citet{2010ApJ...721..715D} find M$_{\rm BH} = (1.7 \pm 0.5) \times 10^{6}$ \Msun\ for NGC 4051 raising the initial estimate but still within the low mass regime.  Low mass AGNs like NGC 4395 and NGC 4051 are useful for constraining evolutionary models since they may challenge in different ways the current coevolution paradigm \citep{2011Natur.469..374K}.    

The robust RM measurements from the moderate luminosity AGNs and the scaling relations derived from it, like the \rblr$-$L relation, can be directly applied to single spectra to determine M$_{\rm BH}$ for large samples of AGNs.  For this so called `single epoch' method, M$_{\rm BH}$ can be calculated using only 2 measurements, the $\Delta V$ of an emission line and the continuum luminosity.  For lower redshift samples $\Delta V$ is usually found using the H$\beta$ line and the luminosity is measured at 5100 \AA\ ($\lambda$L$_{5100}$).  For a review of this method and a comparison with RM measurements see \citet{2009ApJ...692..246D}, where it is shown that for the sufficiently high S/N spectra, the errors associated with single epoch measurements are on the order of 0.1 dex \citep[see also][who estimate errors in M$_{\rm BH}$ from optical and UV scaling relations to be a factor of 4]{2006ApJ...641..689V}.  This method has been increasingly employed in studies of AGNs due to the public availability of large samples of high quality spectra from surveys like the Sloan Digital Sky Survey (SDSS).  Since the slope of the power-law for the \rblr$-$L relation has not been robustly checked at the lowest luminosities, applying the single epoch method to large samples of AGN is an easy way to find low mass candidates for follow-up studies using RM.  The sample of \citet{2004ApJ...610..722G} employs this method to identify 19 candidate AGNs from the SDSS with M$_{\rm BH} < 10^{6}$ \Msun .  From this sample we take SDSS J114008.71+030711.4 (hereafter GH08) for a follow-up reverberation mapping campaign in an attempt to constrain the properties of AGNs hosting the lowest mass BHs.  GH08 has a redshift z $= 0.0811$, which is actually the median of the entire sample.  The SDSS images show a resolved spiral galaxy with a moderately luminous AGN.  The SDSS spectrum of GH08 has fairly narrow broad lines typical of narrow line Seyfert 1 (NLS1) type AGN, which are typically thought to have lower M$_{\rm BH}$ but are accreting at close to the Eddington limit \citep[for a review see][]{2000NewAR..44..455G}.  While line ratios as measured from the SDSS spectrum would place GH08 in the HII region of the BPT diagram \citep[see for example][]{2006MNRAS.372..961K}, there are several detections of an X-ray point source in both ROSAT and {\it Chandra} \citep{2007ApJ...656...84G} at the AGN location.  HST images presented in \citet{2008ApJ...688..159G} reveal GH08 to have a resolved disk component with spiral arms and an elongated bar.  The surface brightness as a function of radius is fit with a three component model (disk, bar and an AGN) and the AGN luminosity after deconvolution from the host is estimated to be $\lambda$L$_{5100}$ $= 5.9 \times 10^{42}$ \ergs\ .  Using the \rblr$-$L relation from \citet[][their equation 2]{2009ApJ...697..160B} and the AGN luminosity estimated from the HST images, we expect GH08 to have \rblr\ $\approx$ 8.7 days. 

In Section 2 we describe the observations, in Section 3 the data analysis, and in Section 4 the conclusions.  We assume the cosmology used by \citet{2004ApJ...610..722G} (H$_{\rm o} = 100$ h $= 72$ \Hubble, $\Omega_{m} = 0.3$, $\Omega_{\lambda} = 0.7$) to compare with their results.

\section{Observations \& Data Reduction}
\indent

The spectral data for GH08 were taken using the 9.2-m Hobby-Eberly Telescope (HET) at McDonald Observatory from December 7, 2009 to June 3, 2010 with a total of 35 observations over a period of 178 days.  The average time between observations is 6.1 days with a median of 5.9.  There are 5 instances where the time between useful observations is less than 3 days.  

All observations were taken using the Low Resolution Spectrograph (LRS) with the g2 grism which has a resolving power of R $=$ 1300 and covers a wavelength region from 4300 $-$ 7300 \AA.  Of these 35 observations, 5 had very low signal to noise (S/N $<$ 2) due to bad observing conditions and are not used in the time series analysis.  All observations have an exposure time of 25 min (providing 30 spectra each with a S/N $\ge$3).  A fixed slit width of 2$''$ was used for each observation, but the position angle of the slit varies randomly from one observation to the next.  A fixed slit position was requested for these queue observations but not obtained.  The errors associated with a changing position angle will be discussed in detail below.

The spectra were reduced using the standard IRAF routines with the appropriate parameters set for the HET LRS instrument.  We use an extraction window of 2.5$''$ to maximize the S/N while minimizing host contamination from scales larger than about 2 kpc.  The spectra were initially flux calibrated using spectral standard stars taken nightly.  On six nights no calibration star was observed, usually due to degrading weather conditions, and the standard star used for the flux calibration of GH08 on those nights was the one observed closest in time to that observation.  Wavelength calibration was done using a HgCdAr and Ne spectra taken nightly.  

To obtain internal flux calibration between the spectra from different nights, the chi-square fitting routine of \citet{1992PASP..104..700V} was employed. In this algorithm the assumption is made that the narrow line flux of the [O{\sc iii}]$\lambda$5007 line is constant, since the NLR is at a large enough distance that the changing continuum will not effect its line flux on these short time scales.  Each individual spectrum was then scaled so that the [O{\sc iii}]$\lambda$5007 line in each spectrum had the same integrated flux as the mean spectrum (from averaging individual spectra) so that calibrated H$\beta$ measurements could be made.  In principle, this procedure should have been repeated using the narrow [S{\sc ii}]$\lambda\lambda$6716,6731 lines to properly calibrate the H$\alpha$ measurements, but the lines are severely blended and the S/N falls drastically just redward of H$\alpha$ so this was not feasible.  Therefore, the results based on the H$\alpha$ line are not as reliable as those for H$\beta$.  Other lines originating in the BLR like He{\sc i}$\lambda$5876 and He{\sc ii}$\lambda$4868 are not strong enough (or absent) given the S/N to make reliable measurements.

The rest frame mean spectrum is shown in the top panel of Figure 1 and was obtained using the highest S/N spectra throughout the run.  The most notable features in the mean spectrum are the strong Balmer lines, the He{\sc i}$\lambda$5876 line, the [O{\sc iii}]$\lambda$4959,5007 doublet and FeII lines flanking H$\beta$.  In the mean spectrum we measure the integrated flux of the [O{\sc iii}]$\lambda$5007 line using an extraction region of 80 \AA\ (rest frame) about the line centroid and find F$_{5007} =$ 1.67 $\times 10^{-15} \ergcms$.  The bottom panel of Figure 1 shows the rest frame RMS spectrum of GH08 after the scaling procedure was completed.  The [O{\sc iii}] doublet disappears as expected and the residual in the H$\beta$ line is the reverberation signal that is to be measured.  The strong residuals in the H$\alpha$, [N{\sc ii}]$\lambda\lambda$6548,6584, and [S{\sc ii}] lines are due to the poorer flux calibration in the red part of the spectrum mentioned above.

\section{Data Analysis}
\indent

The top panel in Figure 2 shows the light curves for the continuum at 5100 \AA ~(rest frame).  To measure the continuum we take a 140 \AA ~wide window centered at 5100 \AA.  This window is just redward of the [O{\sc iii}]$\lambda$5007 line and just blueward of the Fe{\sc ii} lines that start at $\sim$5200 \AA ~(rest frame).  The light curves of the integrated line flux for H$\alpha$ and H$\beta$ are shown in the two bottom panels.  The measured fluxes for the continuum and Balmer lines are listed in Table 1.  The light curve statistics are presented in Table 2.  The three light curves all have a similar global shape, where the variations in the lines are very similar to the variations in the continuum.  They are however different enough from each other that we have confidence that we are not just sampling noise in the seeing conditions or instrument.  There is roughly a factor of 2.3 overall change in the H$\beta$ and H$\alpha$ flux, and nearly a factor of 2 change for the continuum flux (before subtraction of the constant contribution from the host) as shown by R$_{\rm max}$ in Table 2.  A possible source of contamination we consider, due to the observing procedure used, is that the position angle of the slit was not constant throughout the observations.  This contamination can manifest as additional narrow line flux from extended regions when more of the host galaxy is in the slit, thereby changing the relative flux scaling obtained during the inter calibration procedure.  Using images of the slit position with respect to the host galaxy orientation, we have checked that measured changes in flux are not correlated with the slit position angle (thus having more or less of the host galaxy in the slit) at different times throughout the observing run.

For the time series analysis we use two independent methods to cross-correlate the continuum and line light curves.  The first method is the interpolated cross-correlation function (ICCF) of \citet{1986ApJ...305..175G} and \citet{1987ApJS...65....1G}, as implemented by \citet{1994PASP..106..879W}; see also the review by \citet{1994ASPC...69..111G}. The second method is the \emph{z}-transformed discrete correlation function (ZDCF) of \citet{1997ASSL..218..163A} which is an improvement on the discrete correlation function \citep[DCF;][]{1988ApJ...333..646E}. The ZDCF applies Fisher's \emph{z} transformation to the correlation coefficients, and uses equal population bins rather than the equal time bins used in the DCF.  In Figure 3 we show the ICCF and ZDCF for both the H$\alpha$ and H$\beta$ lines.  The value of the ICCF coefficient for H$\beta$ at the central peak is a modest 0.5 and with a time lag of $\sim$2 days, with an error of $+$4 and $-2$ days (observed frame).  The errors in the time lag determination are estimated using the model-independent flux randomization/random subset selection (FR/RSS) Monte Carlo method \citep[for more details see][]{1998PASP..110..660P}.  Our measured value of 2 days is consistent with zero time lag within the given errors, meaning that we need finer time sampling to truly resolve the reverberation signal, but our measurements constitute an upper limit to the time lag.  The time lag of zero for H$\alpha$ could constitute a true lower limit or be spurious due to the poor flux calibration as discussed above.

In Figure 4 we show the velocity profiles of H$\alpha$ and H$\beta$ for the mean spectra along with the Gaussian line fits used to measure the FWHM.  In order to constrain the width of the narrow lines we take the [O{\sc iii}]$\lambda$5007 line as a template and a Gaussian fit gives a FWHM of 519 \kms\ (the effect of instrumental broadening on the line widths and uncertainties is small).  Comparing this to a FWHM of 297 \kms\ as determined by \citet{2004ApJ...610..722G} shows that we are only just resolving the narrow lines.  In order to constrain the contribution of the narrow Balmer lines to the total line profiles we first assume that the [O{\sc iii}]/H$\beta$ flux ratio should be at least 3 based on emission line diagnostics for the classification of type 2 AGN as in \citet{2006MNRAS.372..961K}.  Further, the contribution of the narrow H$\alpha$ component should be roughly 3 times that in the limit of case B recombination.  For H$\alpha$ (top panel) we fit 3 narrow line components and one broad.  All 3 narrow lines are constrained to have the same line widths as [O{\sc iii}].  The narrow [N{\sc ii}] lines are forced to have a flux ratio of 2.96 and the centroids of each line are fixed to the rest frame wavelengths of $\lambda\lambda$6548,6584.  The fit falls below the line profile near the peak. This is due to the fact that a simple Gaussian function is not quite adequate.  The broad component fit should be taller to compensate, which would decrease the FWHM.  Therefore the FWHM of 737 $\pm$ 155 \kms\ we measure for H$\alpha$  can be thought of as an upper limit.  The H$\beta$ line (lower panel) is fit with 2 Gaussians, a narrow component whose width is modeled after the [O{\sc iii}] line and one broad component.  Again the fit falls below the line profile near the peak.  We find an upper limit to the FWHM of H$\beta$ to be 777 $\pm$ 110 \kms.  For NLS1$-$like AGNs a Lorentzian profile can provide a better fit to the wings of the broad component as expected from the analysis of \citet{2011Natur.470..366K} who show that the broad line shapes can vary from more Lorentzian profiles at small FWHM ($\le 2000$ \kms) to more rectangular profiles at larger FWHM ($\ge$ 6000 \kms).  For completeness (but not shown in Figure 4) we also fit both lines with a single Lorentzian assuming that the narrow line contribution is negligible.  Because they are less constrained, the Lorentzians fit better to the overall line profiles.  In table 3 we show the values for each line as measured from the fits as well as the values measured by \citet{2004ApJ...610..722G}.  For the single Lorentzian fits we find better agreement with original measurements of \citet{2004ApJ...610..722G}.  The H$\beta$ line has larger measured FWHM than H$\alpha$ in all cases.  The FWHM of both the H$\beta$ and H$\alpha$ lines are always larger in the mean spectrum than in the RMS spectrum.  This is not surprising given the results of \citet{2006A&A...456...75C} who show that the H$\beta$ line widths are usually 20\% broader in the mean spectrum compared to the RMS spectrum.  We can attribute the overly large widths measured to the fact that the lines are just resolved.  Therefore, since the true line profiles are not expected to be only purely Gaussian or Lorentzian, when estimating the upper limit to M$_{\rm BH}$ we use an H$\beta$ FWHM of 703 $\pm$ 110 \kms\ which is the average of our two H$\beta$ measurements.   

In order to quantify the contribution of the host galaxy to the AGN luminosity we use the analysis by \citet{}. They used HST/ACS images in two filters (B band $-$ F435W and I band $-$ F814W) to decompose the observed intensity distribution of all the images of the \citet{2004ApJ...610..722G} sample into a combination of components, which includes a central unresolved source and some combination of a possible disk, bulge, and a bar.  In the case of GH08 they do not fit for a bulge, possibly due to the contrast with the strong bar.  However, the presence of a small bulge can not be ruled out from a visual inspection of the images.  \citet{2008ApJ...688..159G} plot in their Figure 1 the contributions of the AGN, the bar, and the disk components as a function of radius, as well as the total surface brightness in the I band and the B$-$I color.  We use these components to estimate the AGN contribution to the total flux within a circular aperture of radius 2$''$ by integration over the surface brightness profiles plotted for the different components of GH08. We find that in the I band the AGN contribution is about 20\% of the total flux and in the B band it is about 50\% of the total flux.  Using the GH08 AGN luminosity given in \citet{2008ApJ...688..159G} and the HST images analysis we described above, we find the galaxy contribution to the luminosity in the central 2$''$ of GH08 to be $\lambda$L$_{5100} = 7.5 \times 10^{42}$ \ergs (F$_{5100} \approx 1.1 \times 10^{-16}$ \ergcmsA).  Subtracting this contribution from the total luminosity we measure in this study, we estimate the AGN luminosity at 5100 \AA\ to be about $\lambda$L$_{5100} = 1.1 \times 10^{43}$ \ergs (F$_{5100} \approx 1.6 \times 10^{-16}$ \ergcmsA).

We also used the SDSS images of GH08 to estimate the galaxy contribution.  We used the IRAF isophot package to fit ellipses to the SDSS images of the object and linearly extrapolate the radial profile of the elliptical isophotes intensities in the outer 2$''-$5$''$ of the image to the inner 2$''$ (same as used above with the HST data).  The ratio of counts of the measured profile inside 2$''$ to the extrapolation from outside 2$''$ is taken as a measure of the AGN fraction with respect to the host. We find the AGN to be 46\% and 36\% of the total luminosity in the SDSS $g`$ and $r`$ bands respectively, and we take the average of these two, 41\%, as the AGN fraction.  We note that the AGN fraction may be closer to 46\% since the continuum region at 5100 \AA ~is close to the red side of the SDSS $g`$ band.  From this extrapolation method we find for the AGN that $\lambda$L$_{5100} \approx 7.6 \times 10^{42}$ \ergs (F$_{5100} \approx 1.1 \times 10^{-16}$ \ergcmsA) with an upper limit of $1.8 \times 10^{43}$ \ergs (F$_{5100} = 2.6 \times 10^{-16}$ \ergcmsA).  Given the amplitude of continuum variability as shown in Figure 2, these results are in good agreement with the results we have calculated above from the HST/ACS images analysis done by \citet{2008ApJ...688..159G}.  In the following we take the AGN luminosity at 5100 \AA\ to be $1.1 \times 10^{43}$ \ergs\ with an an upper limit of $1.8 \times 10^{43}$ \ergs .

Combining all the above measurements and using Equation (1) we find for GH08 that M$_{\rm BH} = 1.5 \times 10^{5}$ \Msun with an upper limit of M$_{\rm BH} \lesssim 5.8 \times 10^{5}$ \Msun .  The bolometric luminosity, if L$_{\rm bol} \approx$ 10L$_{\rm 5100}$ \citep[][]{1994ApJS...95....1E, 2004MNRAS.352.1390M,2006ApJS..166..470R}, implies L$_{\rm bol}/$L$_{\rm Edd} = 1.5$.  In Figure 5 we show GH08 in the \rblr$-$L and M$_{\rm BH} -$L planes (triangle), with additional sources (circles) compiled from \citet{2006ApJ...644..133B}, \citet{2009ApJ...697..160B}, \citet{2009ApJ...705..199B} and \citet{2010ApJ...721..715D}.  The position of GH08 on the \rblr$-$L plot is just within the observed scatter.  GH08 deviates more strongly in the M$_{\rm BH} -$L plot.  With the upper limits on both L$_{\rm 5100}$ and M$_{\rm BH}$ GH08 is even more outlying.  This is due to the high accretion rate and explains the moderate luminosity for such a low M$_{\rm BH}$ AGN, as indicated by the lines of constant Eddington ratio in the bottom panel of Figure 5.

We can compare the presently determined M$_{\rm BH}$ with that expected from the M$-\sigma_{*}$ relation.  This will give insight into the the coevolution (or lack thereof) of the bulge in AGN with low M$_{\rm BH}$.  \citet{2005ApJ...619L.151B} measure $\sigma_{*}$ for 15 of the \citet{2004ApJ...610..722G} objects (but not GH08) from high resolution spectra and find that most of the objects fit well the extrapolation to lower M$_{\rm BH}$ using the M$-\sigma_{*}$ relation for more massive BHs.  As these authors mention in their discussion, uncertainties in the analysis rely moderately on M$_{\rm BH}$ determined using the single epoch method and the selection criteria, where the more luminous, and hence the more massive BHs are selected.  For our analysis we loosely assume that $\sigma_{\rm [O{\sc III}]} \propto \sigma_{*}$ \citep[see e.g.,][]{1996ApJ...465...96N, 2006ApJ...636..654R}.  Using the relation given by \citet{2009ApJ...698..198G} and a measured $\sigma_{\rm [O{\sc III}]} =$ 126 \kms\ (derived from \citet{2004ApJ...610..722G} who measure FWHM$_{\rm [O{\sc III}]} =$ 297 \kms ) we would expect M$_{\rm BH,*} \approx 1.9 \times 10^{7}$ \Msun.  The surprising discrepancy with the measured M$_{\rm BH}$ which is as large as $\sim$1.5 dex probably reflects the large uncertainty associated with using the narrow [O{\sc iii}] line as a proxy for $\sigma_{*}$, as it has been shown that asymmetry in the blue wings and systematic blue shifts of the [O{\sc iii}] line centroid are observed for many of the NLS1-like objects \citep{2005AJ....130..381B, 2007ApJ...667L..33K}.  Also, a $\sigma_{*} = 127$ \kms\ for GH08 is about twice the average measured $\sigma_{*}$ in the \citet{2005ApJ...619L.151B} sample so it is likely that this value is an overestimation but the discrepancy from the RM results will still persist.  It is still a point of debate as to whether or not AGNs with low M$_{\rm BH}$ and NLS1 objects like GH08 genuinely deviate from (lie below) the M$-\sigma_{*}$ relation \citep[for more in depth discussions see][]{2008ApJ...688..159G, 2011arXiv1102.0537M, 2011arXiv1106.6232X}.  This leaves one to speculate on the true impact AGNs with intermediate BHs have in terms of feedback on the host properties and the AGNs ability to regulate simultaneous growth of a bulge.

\section{Conclusion}
\indent

We attempt to measure the central M$_{\rm BH}$ in the intermediate mass AGN candidate GH08 using the RM technique.  GH08 is originally from the sample of \citet{2004ApJ...610..722G} and is a low redshift, late type spiral galaxy and hosts a moderate luminosity AGN of the NLS1 variety.  Our spectroscopic monitoring campaign spans 178 days yielding 30 useful observations.  From these data we measure the H$\beta$ time lag to be 2 days, with an upper limit of 6 days.  To further constrain and robustly resolve the time lag will require much denser time sampling for this object.  Combining the measured $\tau$ with the H$\beta$ FWHM of 800 \kms, we find M$_{\rm BH} = 1.5 \times 10^{5}$ \Msun\ with an upper limit of M$_{\rm BH} \leq 5.8 \times 10^{5}$ \Msun .  This may constitute the lowest M$_{\rm BH}$ and highest L/L$_{\rm Edd}$ AGN measured from ground based observations by RM to date.  GH08 deviates from other low mass AGN candidates like NGC 4395, which notoriously lacks a bulge, in that it has bar and therefore likely has substantial vertical structure above the galactic disk (most likely supported near the center by velocity dispersion), and therefore is more likely to follow the M$-\sigma_{*}$ relation, which predicts a larger BH mass than the one we measure.  While relations like M$-\sigma_{*}$ provide evidence for AGN feedback and their impact on galaxy evolution, further investigation into low M$_{\rm BH}$ objects like GH08 will constrain deviations from the current paradigm for the coevolution of the central SMBH and the host galaxy.

\acknowledgments
This work has been supported by the Niedersachsen-Israel Research Cooperation Program ZN2318.  S.K. is supported at the Technion by the Kitzman Fellowship and by a grant from the Israel-Niedersachsen collaboration program.  S.R. is supported at the Technion by the Lady Davis Fellowship.  Based on observations obtained with the Hobby-Eberly Telescope, which is a joint project of the University of Texas at Austin, the Pennsylvania State University, Stanford University, Ludwig-Maximilians-Universit\"at M\"unchen, and Georg-August-Universit\"at G\"ottingen.  We acknowledge useful discussions with Ari Laor and Jonathan Stern as well as helpful comments and suggestions from Jenny Greene, Misty Bentz and Ohad Shemmer.

\newpage

\begin{figure}[t]
\includegraphics[angle=0, scale=0.7]{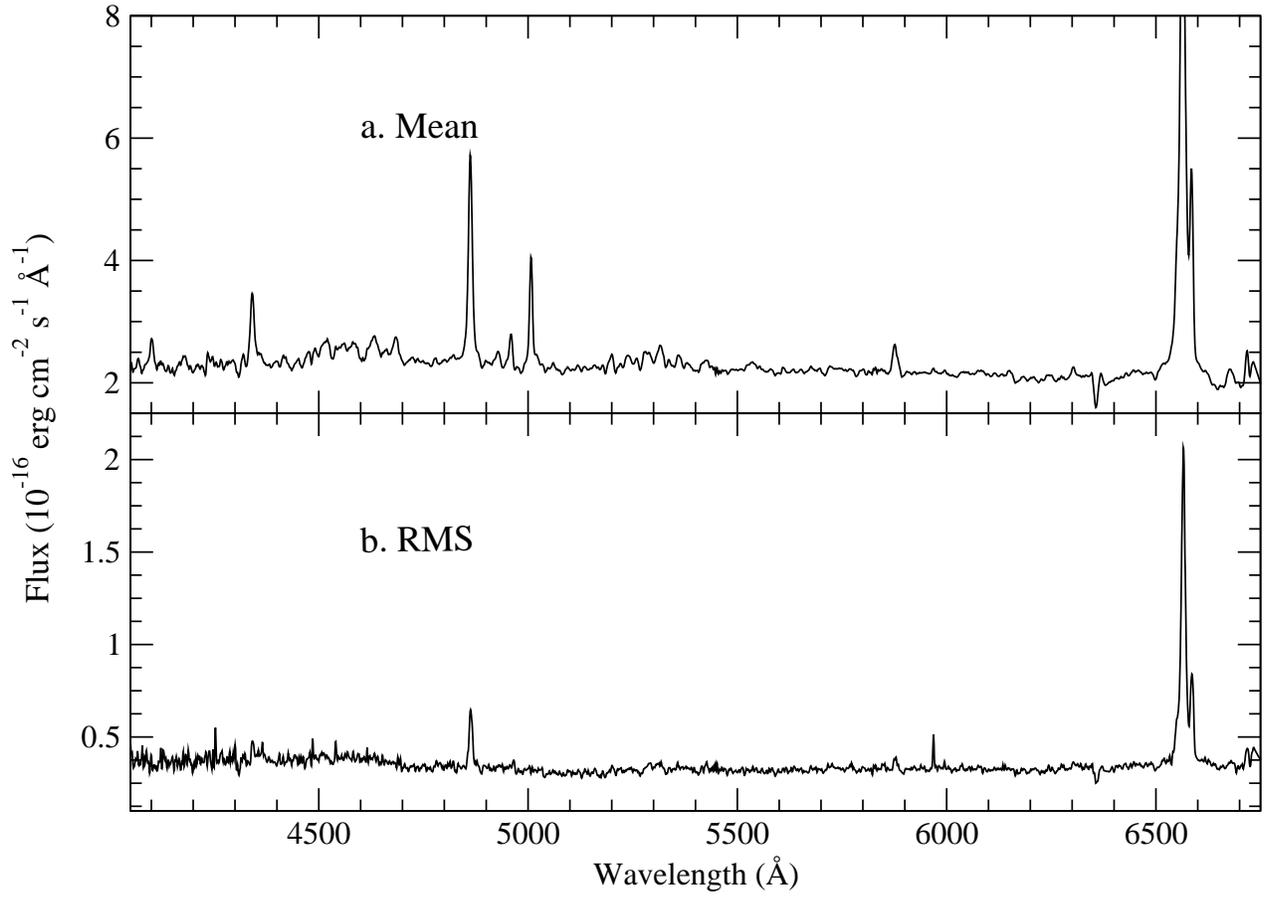}
\caption[]{a. Mean spectrum of GH08; b. RMS spectrum of GH08.}
\label{fig1}
\end{figure}

\begin{figure}[b]
\includegraphics[angle=0, scale=0.7]{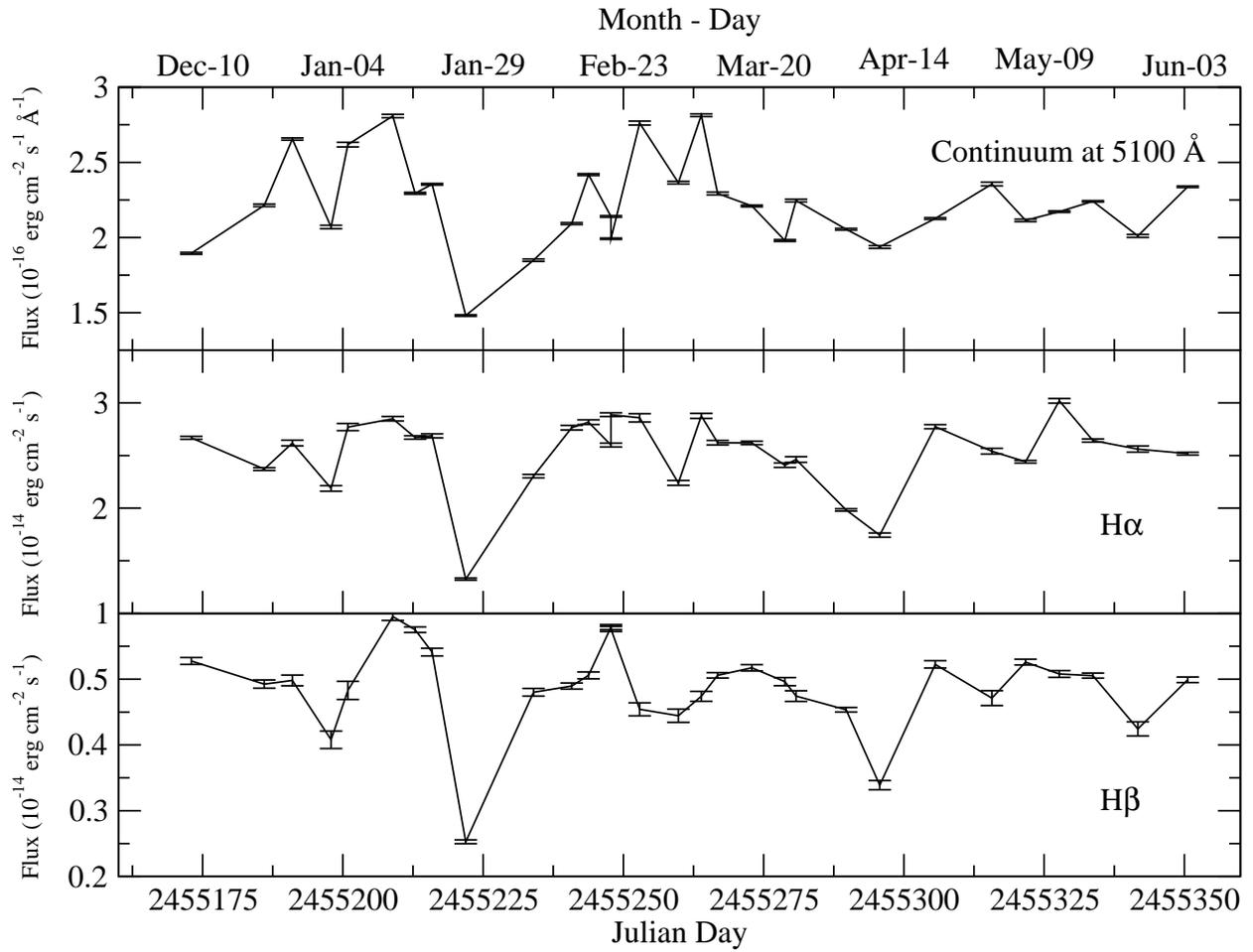}
\caption[]{Light curves for the continuum at 5100 \AA (top), H$\alpha$ (middle) and H$\beta$ (bottom).}
\label{fig2}
\end{figure}

\begin{figure}[b]
\includegraphics[angle=0, scale=0.7]{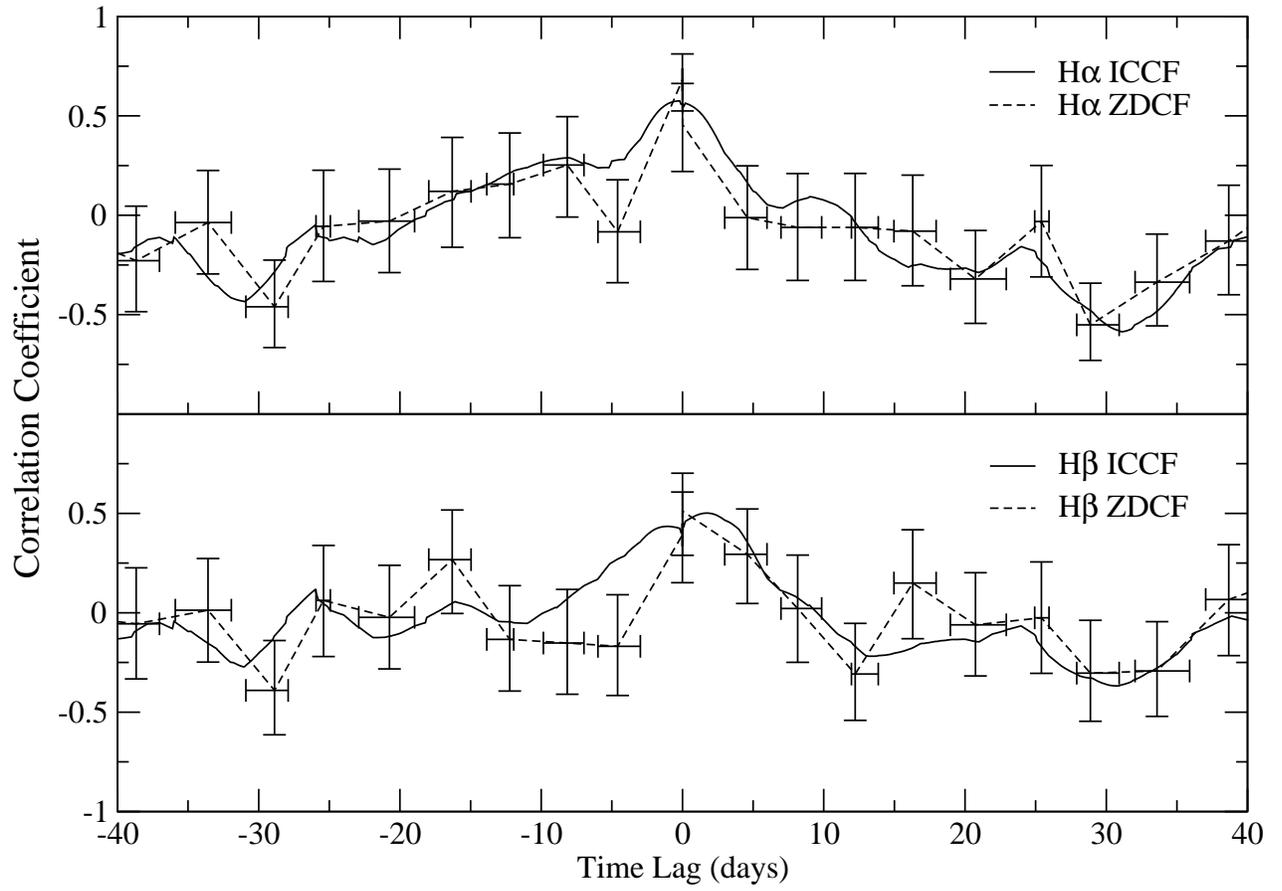}
\caption[]{Solid line is the ICCF and the dashed line is the ZDCF; Top: H$\alpha$ CCF; Bottom: H$\beta$ CCF.}
\label{fig3}
\end{figure}

\begin{figure}[b]
\includegraphics[angle=0, scale=0.7]{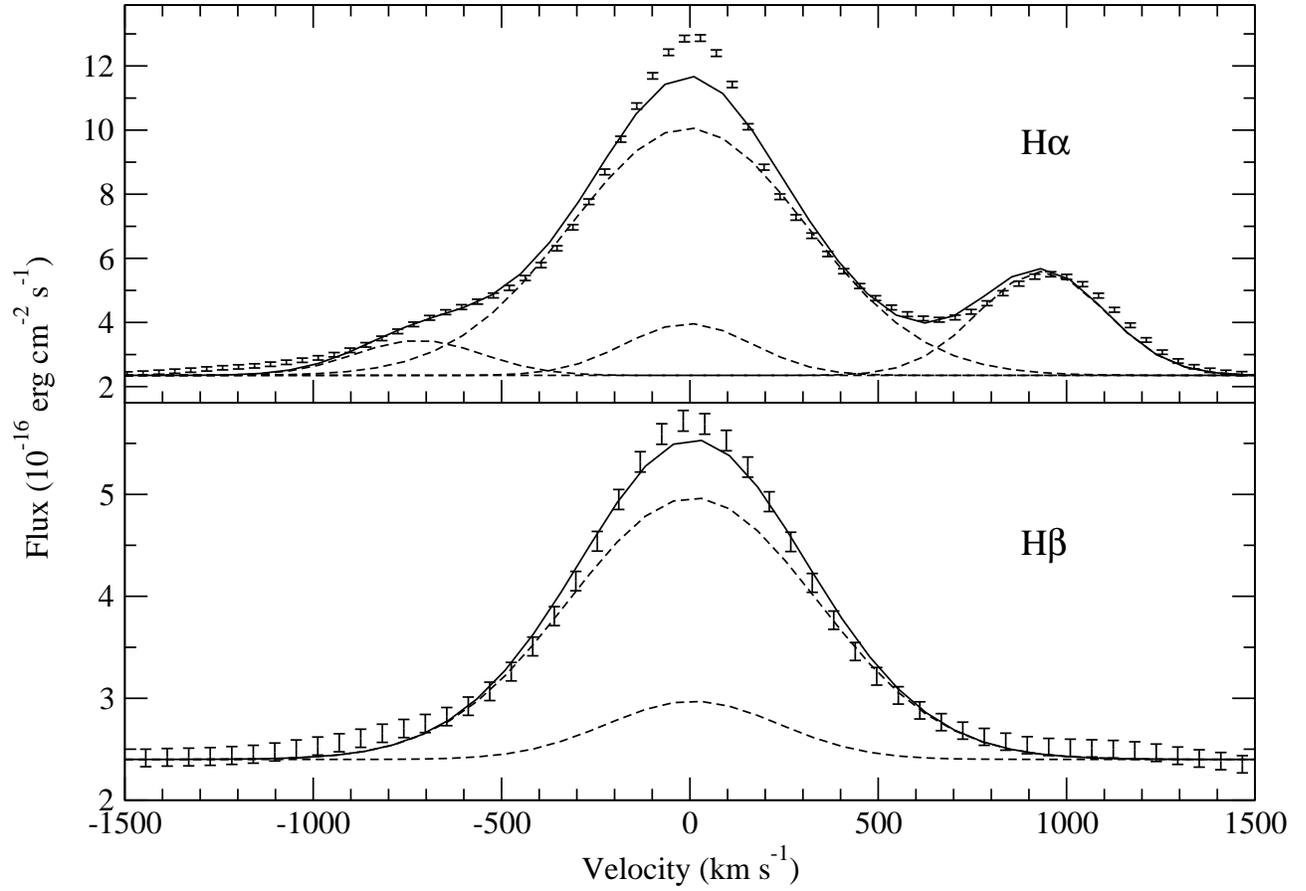}
\caption[]{Top: The black data points are the H$\alpha$ mean velocity profile, the dotted lines are the Gaussian fit for each component, and the solid line is the total fit to the line profile.  Bottom: Same as above but for  the H$\beta$ mean velocity profile.  Both fits fall short at the peak due to fitting with pure Gaussians, therefore the measured FWHM can be considered to be upper limits.  Measured values are presented in Table 3.}
\label{fig4}
\end{figure}

\begin{figure}[b]
\includegraphics[angle=0, scale=0.7]{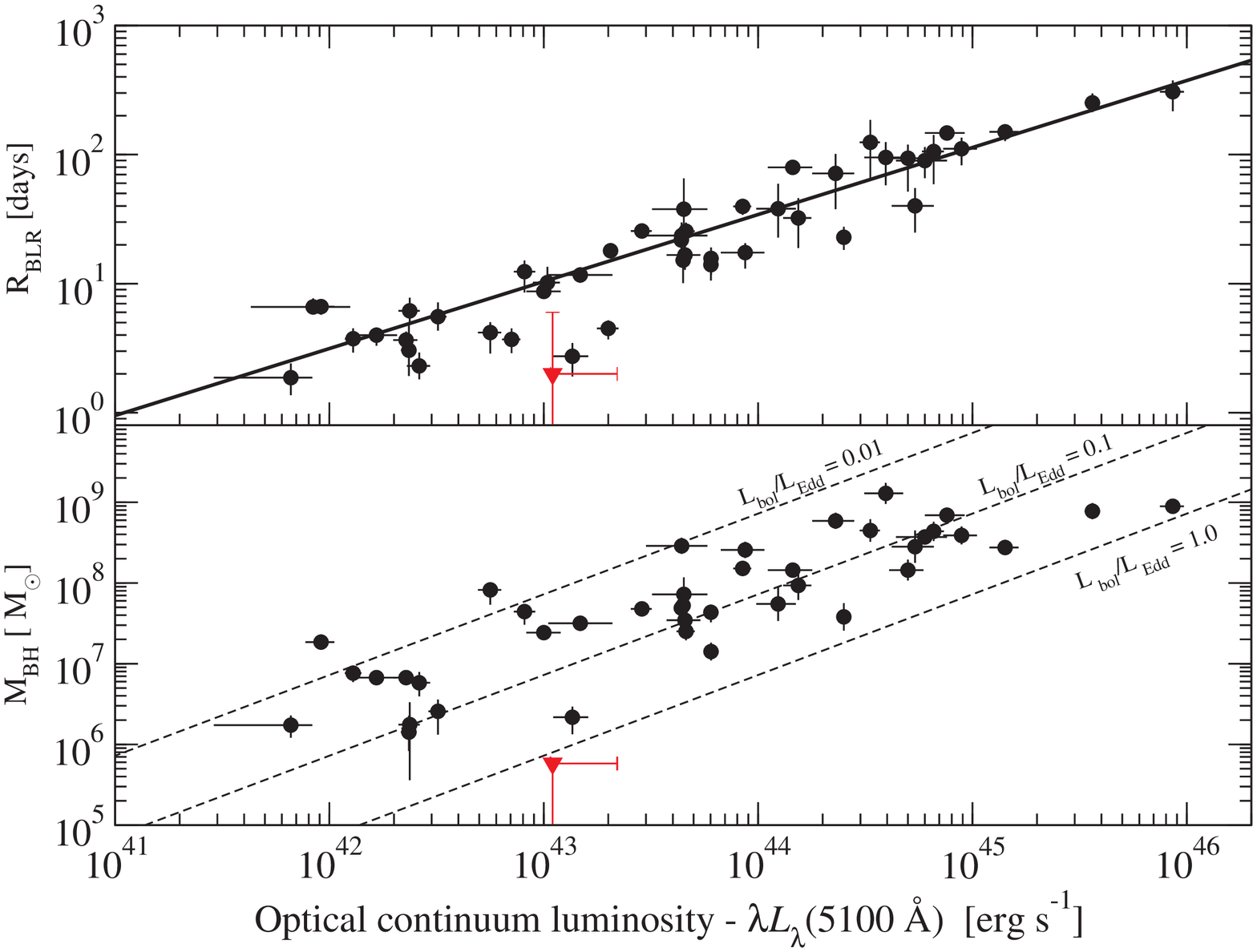}
\caption[]{Top: R$_{\rm BLR}$ vs. L$_{\rm 5100}$.  Data points (all measured using H$\beta$ time lag) are from \citet{2009ApJ...697..160B}, \citet{2009ApJ...705..199B}, and \citet{2010ApJ...721..715D}.  The solid line is the current best published estimate for the R$_{\rm BLR} - $L relation by \citet{2009ApJ...697..160B} (it is not a fit to all the shown data points -- see \citet{2010ApJ...721..715D} for details); Bottom: M$_{\rm BH}$ vs. L$_{\rm 5100}$. Data points are from \citet{2009ApJ...697..160B}, \citet{2009ApJ...705..199B}, and \citet{2010ApJ...721..715D}.  The dashed lines are lines of constant Eddington ratio (L$_{\rm bol}/$L$_{\rm Edd} =$ 0.01, 0.1, 1.0 from the top).  In both panels the point of GH08 is shown as a red triangle.}
\label{fig5}
\end{figure}

\begin{deluxetable}{lccc}
\tablecolumns{4}
\tabletypesize{\footnotesize}
\tablewidth{0pc}
\tablecaption{Continuum, H$\alpha$ and H$\beta$ Light Curves\label{table}}
\tablehead{
	\colhead{Julian Date} & 
	\colhead{F$_{\rm 5100 \AA}$} & 
	\colhead{F$_{\rm H\alpha}$} &
	\colhead{F$_{\rm H\beta}$} \\
	\colhead{} & 
	\colhead{} & 
	\colhead{} & 
	\colhead{} \\
\colhead{(1)} & \colhead{(2)} & \colhead{(3)} & \colhead{(4)}
}
\startdata
5172.99 & 1.895 $\pm$ 0.007 & 2.668 $\pm$ 0.014 & 0.528 $\pm$ 0.005 \\
5185.96 & 2.214 $\pm$ 0.008 & 2.371 $\pm$ 0.013 & 0.492 $\pm$ 0.006 \\
5190.96 & 2.655 $\pm$ 0.008 & 2.619 $\pm$ 0.027 & 0.498 $\pm$ 0.008 \\
5197.93 & 2.070 $\pm$ 0.011 & 2.187 $\pm$ 0.026 & 0.408 $\pm$ 0.013 \\
5200.92 & 2.618 $\pm$ 0.015 & 2.770 $\pm$ 0.034 & 0.483 $\pm$ 0.014 \\
5208.91 & 2.808 $\pm$ 0.012 & 2.849 $\pm$ 0.021 & 0.595 $\pm$ 0.006 \\
5212.89 & 2.294 $\pm$ 0.006 & 2.671 $\pm$ 0.016 & 0.575 $\pm$ 0.004 \\
5215.88 & 2.354 $\pm$ 0.005 & 2.687 $\pm$ 0.019 & 0.541 $\pm$ 0.006 \\
5221.86 & 1.481 $\pm$ 0.005 & 1.326 $\pm$ 0.010 & 0.253 $\pm$ 0.003 \\
5233.95 & 1.849 $\pm$ 0.008 & 2.304 $\pm$ 0.017 & 0.480 $\pm$ 0.006 \\
5240.82 & 2.093 $\pm$ 0.007 & 2.764 $\pm$ 0.022 & 0.490 $\pm$ 0.005 \\
5243.80 & 2.419 $\pm$ 0.006 & 2.817 $\pm$ 0.022 & 0.506 $\pm$ 0.005 \\
5247.79 & 2.140 $\pm$ 0.005 & 2.600 $\pm$ 0.018 & 0.579 $\pm$ 0.004 \\
5247.81 & 1.992 $\pm$ 0.005 & 2.888 $\pm$ 0.018 & 0.576 $\pm$ 0.004 \\
5252.91 & 2.761 $\pm$ 0.013 & 2.858 $\pm$ 0.039 & 0.454 $\pm$ 0.010 \\
5259.76 & 2.365 $\pm$ 0.008 & 2.240 $\pm$ 0.023 & 0.444 $\pm$ 0.010 \\
5263.88 & 2.813 $\pm$ 0.009 & 2.876 $\pm$ 0.024 & 0.474 $\pm$ 0.007 \\
5266.77 & 2.293 $\pm$ 0.010 & 2.622 $\pm$ 0.022 & 0.506 $\pm$ 0.004 \\
5272.87 & 2.210 $\pm$ 0.006 & 2.620 $\pm$ 0.016 & 0.517 $\pm$ 0.005 \\
5278.84 & 1.980 $\pm$ 0.006 & 2.408 $\pm$ 0.020 & 0.496 $\pm$ 0.006 \\
5280.82 & 2.246 $\pm$ 0.009 & 2.462 $\pm$ 0.027 & 0.474 $\pm$ 0.008 \\
5289.68 & 2.055 $\pm$ 0.006 & 1.983 $\pm$ 0.011 & 0.454 $\pm$ 0.003 \\
5295.66 & 1.937 $\pm$ 0.009 & 1.744 $\pm$ 0.020 & 0.339 $\pm$ 0.007 \\
5305.64 & 2.126 $\pm$ 0.006 & 2.774 $\pm$ 0.019 & 0.523 $\pm$ 0.006 \\
5315.74 & 2.356 $\pm$ 0.012 & 2.541 $\pm$ 0.028 & 0.471 $\pm$ 0.011 \\
5321.72 & 2.114 $\pm$ 0.008 & 2.441 $\pm$ 0.013 & 0.526 $\pm$ 0.004 \\
5327.72 & 2.172 $\pm$ 0.006 & 3.020 $\pm$ 0.022 & 0.508 $\pm$ 0.005 \\
5333.68 & 2.241 $\pm$ 0.005 & 2.642 $\pm$ 0.015 & 0.505 $\pm$ 0.004 \\
5341.66 & 2.011 $\pm$ 0.010 & 2.561 $\pm$ 0.029 & 0.424 $\pm$ 0.011 \\
5350.64 & 2.338 $\pm$ 0.005 & 2.517 $\pm$ 0.014 & 0.499 $\pm$ 0.004 \\
\enddata
\tablecomments{
Col.(1): J.D. $-$2450000 days;
Col.(2): Continuum flux (10$^{-16}$ erg cm$^{-2}$ s$^{-1}$ \AA$^{-1}$);
Col.(3) \& (4): Line flux (10$^{-14}$ erg cm$^{-2}$ s$^{-1}$).  The length of each exposure is 0.0174 days.
}
\end{deluxetable}

\clearpage

\begin{deluxetable}{ccccccc}
\tablecolumns{7}
\tabletypesize{\footnotesize}
\tablewidth{0pc}
\tablecaption{Continuum, H$\alpha$ and H$\beta$ Light Curve Statistics\label{table}}
\tablehead{
	\colhead{Cont./Line} & 
	\colhead{F$_{\rm min}$} & 
	\colhead{F$_{\rm max}$} &
	\colhead{R$_{\rm max}$} &
	\colhead{F$_{\rm avg}$} & 
	\colhead{$\sigma_{\rm F}$} & 
	\colhead{F$_{\rm var}$} \\
\colhead{(1)} & \colhead{(2)} & \colhead{(3)} & \colhead{(4)}
& \colhead{(5)} & \colhead{(6)} & \colhead{(7)}
}
\startdata
Cont. at 5100 \AA & 1.48 & 2.81 & 1.90 & 2.23 & 0.298 & 0.134 \\
Cont. at 1500 \AA & 0.38 & 1.71 & 4.50 & 1.13 & 0.298 & 0.287 \\
after host subtracted \\ 
H$\beta$ & 0.25 & 0.59 & 2.35 & 0.48 & 0.069 & 0.141 \\
H$\alpha$ & 1.33 & 3.02 & 2.28 & 2.53 & 0.359 & 0.142 \\
\enddata
\tablecomments{
Continuum flux in units of 10$^{-16}$ erg cm$^{-2}$ s$^{-1}$ \AA$^{-1}$;
Line flux in units of 10$^{-14}$ erg cm$^{-2}$ s$^{-1}$; R$_{\rm max} =$ F$_{\rm max}/$F$_{\rm min}$; $\sigma_{\rm F}$ is the standard deviation; F$_{\rm var}$ is the fractional variation as defined by \citet{1997ApJS..110....9R}.
}
\end{deluxetable}
\clearpage

\begin{deluxetable}{cccc}
\tablecolumns{4}
\tabletypesize{\footnotesize}
\tablewidth{0pc}
\tablecaption{[O{\sc iii}], H$\alpha$ and H$\beta$ Measured Velocities\label{table}}
\tablehead{
	\colhead{Line} & 
	\colhead{Gaussian FWHM} & 
	\colhead{Lorentzian FWHM} &
	\colhead{G\&H04} \\
	\colhead{} &
	\colhead{(km s$^{-1}$)} &
	\colhead{(km s$^{-1}$)} &
	\colhead{(km s$^{-1}$)} \\
\colhead{(1)} & \colhead{(2)} & \colhead{(3)} & \colhead{(4)}
}
\startdata
[O{\sc iii}] & 519 $\pm$ 11  & 454 $\pm$ 19 & 297 \\
H$\alpha$    & 737 $\pm$ 155 & 540 $\pm$ 6  & 591 \\
H$\beta$     & 777 $\pm$ 111 & 629 $\pm$ 12 & $-$ \\
\enddata
\tablecomments{~Line widths for [O{\sc iii}], H$\alpha$ and H$\beta$ in km s$^{-1}$.  The Gaussian fits include both a narrow and broad component for the Balmer lines and are shown in Figure 4.  The Lorentzian fits are for a single component where the narrow line contribution is assumed to be negligible.  The values in the final column are from \citet{2004ApJ...610..722G}.
}
\end{deluxetable}
\clearpage

\end{document}